\documentclass[aps,prl,twocolumn,showpacs,amsmath,amssymb]{revtex4-1}
\usepackage{amsmath}
\usepackage{graphicx}
\usepackage{subfigure}
\usepackage{epstopdf}
\usepackage{color}
\usepackage{multirow}
\usepackage{setspace}
\usepackage{overpic}
\usepackage{amssymb}
\usepackage[bookmarksnumbered, pdfstartview=FitH,colorlinks,urlcolor=blue, citecolor=blue,linkcolor=blue] {hyperref}
\usepackage{lineno}
\usepackage{bm}
\usepackage{rotating}
\usepackage[utf8]{inputenc}
\usepackage{ulem}

\let\oldequation\equation
\let\oldendequation\endequation

\renewenvironment{equation}
  {\linenomathNonumbers\oldequation}
  {\oldendequation\endlinenomath}

\begin{document}

\title{\bf \boldmath
Measurements of Absolute Branching Fractions of $D^0\to K_L^0\phi$, $K_L^0\eta$, $K_L^0\omega$, and $K_L^0\eta^{\prime}$}

\author{
M.~Ablikim$^{1}$, M.~N.~Achasov$^{10,c}$, P.~Adlarson$^{67}$, S. ~Ahmed$^{15}$, M.~Albrecht$^{4}$, R.~Aliberti$^{28}$, A.~Amoroso$^{66A,66C}$, M.~R.~An$^{32}$, Q.~An$^{63,49}$, X.~H.~Bai$^{57}$
% ...
\\
\vspace{0.2cm}
(BESIII Collaboration)\\
\vspace{0.2cm} {\it
$^{1}$ Institute of High Energy Physics, Beijing 100049, People's Republic of China\\
% ...
$^{4}$ Bochum Ruhr-University, D-44780 Bochum, Germany\\
% ...
$^{10}$ G.I. Budker Institute of Nuclear Physics SB RAS (BINP), Novosibirsk 630090, Russia\\
% ...
$^{15}$ Helmholtz Institute Mainz, Staudinger Weg 18, D-55099 Mainz, Germany\\
% ...
$^{28}$ Johannes Gutenberg University of Mainz, Johann-Joachim-Becher-Weg 45, D-55099 Mainz, Germany\\
% ...
$^{32}$ Liaoning Normal University, Dalian 116029, People's Republic of China\\
% ...
$^{49}$ State Key Laboratory of Particle Detection and Electronics, Beijing 100049, Hefei 230026, People's Republic of China\\
% ...
$^{57}$ University of Jinan, Jinan 250022, People's Republic of China\\
% ...
$^{63}$ University of Science and Technology of China, Hefei 230026, People's Republic of China\\
% ...
$^{66}$ University of Turin and INFN, (A)University of Turin, I-10125, Turin, Italy; (B)University of Eastern Piedmont, I-15121, Alessandria, Italy; (C)INFN, I-10125, Turin, Italy\\
$^{67}$ Uppsala University, Box 516, SE-75120 Uppsala, Sweden\\
% ...
$^{c}$ Also at the Novosibirsk State University, Novosibirsk, 630090, Russia\\
% ...
}
}

\begin{abstract}
We report the first measurements of the absolute branching fractions of $D^0\to K_L^0\phi$, $D^0\to K_L^0\eta$, $D^0\to K_L^0\omega$, and $D^0\to K_L^0\eta^{\prime}$, by analyzing $2.93\,\rm fb^{-1}$ of $e^+e^-$ collision data taken at a center-of-mass energy of 3.773 GeV with the BESIII detector. Taking the world averages of the branching fractions of $D^0\to K_S^0\phi$, $D^0\to K_S^0\eta$, $D^0\to K_S^0\omega$, and $D^0\to K_S^0\eta^{\prime}$, the $K_S^0$-$K_L^0$ asymmetries $\mathcal{R}(D^0, X)$ in these decay modes are obtained. The $CP$ asymmetries in these decays are also determined. No significant $CP$ violation is observed.
\end{abstract}

\maketitle

\oddsidemargin  -0.2cm
\evensidemargin -0.2cm

Hadronic decays of charmed mesons offer an ideal test-bed to investigate strong and weak interactions.
Remarkable progress in studies of hadronic $D$ decays involving $K^\pm$ and $K_S^0$ has been achieved to date.
However, experimental knowledge of hadronic $D$ decays involving a $K_L^0$ is still very poor~\cite{pdg2020} mainly due to the difficulty in $K^0_L$ reconstruction.
It is often assumed (or taken as a good approximation) that the branching fractions (BFs) of $D$ decays into hadronic final states containing $K_L^0$ meson(s) are equal to those for the corresponding final states with $K_S^0$ meson(s).
However, as clarified in Refs.~\cite{asymmetry_theory1, asymmetry_theory2,Rosner:2006bw, asymmetry_theory3,lanzhou_theory,Yu:2017oky}, the interference between Cabibbo-Favored (CF) and Doubly-Cabibbo-Suppressed (DCS) amplitudes can lead to a significant asymmetry between the BFs of $D^0\to K_S^0X$ and $D^0\to K_L^0X$ ($X=\pi^0$, $\eta$, $\eta^\prime$, $\omega$, $\rho^0$, and $\phi$),
\begin{eqnarray}
{\mathcal R}(D^0,X)&=&\frac{\mathcal{B}(D^0\to K_S^0 X)-\mathcal{B}(D^0\to K_L^0 X)}{\mathcal{B}(D^0\to K_S^0 X)+\mathcal{B}(D^0\to K_L^0 X)}\nonumber \\
                 &=&-2 r \cos\delta +y_D,
\end{eqnarray}
where $r$ and $\delta$ are the relative strength and phase between the DCS and CF amplitudes, respectively, and $y_D$ is the $D^0$-$\bar{D}^0$ mixing parameter~\cite{mixing_parameter}.
One has ${\mathcal R}(D^0,P)=2{\rm tan}^2\theta_C(+y_D)=0.113\pm0.001$ for $P=\pi^0$, $\eta$, or $\eta^\prime$ naively~\cite{asymmetry_theory1,Rosner:2006bw,asymmetry_theory2,asymmetry_theory3,lanzhou_theory},
where $\theta_C$ is the Cabibbo mixing angle~\cite{asy_explain}.
Using the factorization-assisted topological (FAT) amplitude approach and assuming $E_{P}=E_{V}$, Ref.~\cite{lanzhou_theory} stated that
the ${\mathcal R}(D^0,V)$ for $V=\rho$, $\omega$, or $\phi$  can also be simplified as $2{\rm tan}^2\theta_C+y_D=0.113\pm0.001$,
where $E_{P}$ and $E_{V}$ are the $W$-exchange amplitudes for $D\to PP$ and $D\to VP$ decays, respectively.
Here, $P$ and $V$ denote pseudoscalar and vector mesons, respectively.
It is independent of $X$ because the ratio of DCS and CF amplitudes only depends on the Cabibbo-Kobayashi-Maskawa (CKM) matrix elements.
The large asymmetry for ${\mathcal R}(D^0,\pi^0)$ has been confirmed by a previous measurement of the CLEO experiment~\cite{cleo_exp}.
Measurements of the BFs of $D^0\to K_L^0\phi$, $D^0\to K_L^0\eta$, $D^0\to K_L^0\omega$, and $D^0\to K_L^0\eta^{\prime}$ are crucial to test theoretical calculations and help understand the CKM mechanism.
Study of the $K_S^0$-$K_L^0$ asymmetry, $\mathcal{R}(D,X)$ is also important to improve the understanding of
quark U-spin~\cite{Kingsley,Gronau} and SU(3)-flavor symmetry breaking effects and can benefit theoretical predictions of $CP$ violation in $D$ decays~\cite{ref5,theory_a,theory_5,theory_4,theory_3,theory_2,theory_1,zzxing,qqin}.
These decays are all $CP+$ eigenstates and can be used to extract the strong phase differences of neutral $D$ decays~\cite{gamma_angle1, gamma_angle2}.

Studies of $CP$ violation of the weak decays of $D$ mesons are important for exploring physics within and beyond the Standard Model.
The size of $CP$ violation in various $D$ decays is predicted to be in the order of $10^{-3}$~\cite{ref1,ref2,ref3,ref4,ref5,ref6,ref7}.
In 2019, LHCb reported the first observation of $CP$ violation in neutral $D$ decays~\cite{lhcb_D_CP}.
Currently, the knowledge of $CP$ violation in the charm sector is still limited and further measurements are highly desirable. 

This paper reports the first measurements of the BFs of $D^0\to K_L^0\phi$, $D^0\to K_L^0\eta$, $D^0\to K_L^0\omega$, and $D^0\to K_L^0\eta^{\prime}$
as well as the BF asymmetries between $D^{0}\to K_{S}^0X $ and  $D^0\to K_{L}^0X$.
In addition, the $CP$ asymmetries in these decays are also determined.
Throughout this paper, charge conjugate channels are implied, unless noted otherwise.

This analysis is performed with a 2.93 fb$^{-1}$~\cite{lum_bes3} sample of $e^+e^-$ annihilation data taken at a center-of-mass energy $\sqrt s=3.773$~GeV with the BESIII detector.
Details about the design and performance of the BESIII detector are given in Ref.~\cite{BESCol}.
Simulated samples, produced with the {\sc geant4}-based~\cite{geant4} Monte Carlo (MC) package including the geometric description of the BESIII detector and the
detector response, are used to determine the detection efficiency
and to estimate background contributions. The simulation includes the beam-energy spread and initial-state radiation in the $e^+e^-$
annihilations modeled with the generator {\sc kkmc}~\cite{kkmc}.
An inclusive MC sample, containing the production of $D\bar{D}$
pairs, the non-$D\bar{D}$ decays of the $\psi(3770)$, the initial-state radiation
production of the $J/\psi$ and $\psi(3686)$ states, and the
continuum processes, is used in this analysis. Known decay modes are modeled with {\sc
evtgen}~\cite{evtgen} using the BFs taken from the
Particle Data Group (PDG)~\cite{pdg2020} and the remaining unknown decays
from the charmonium states are modeled with {\sc
lundcharm}~\cite{lundcharm}. Final state radiation
from charged final-state particles is incorporated using {\sc
photos}~\cite{photos}.

At $\sqrt s=3.773$~GeV, $D^0$ and $\bar D^0$ mesons are produced in pairs without accompanying hadrons, and hence the environment is ideal to investigate $D^0$ decays with the double-tag (DT) method~\cite{DT_method}.
In this method, the $\bar{D}^0$ meson, later referred as single-tag (ST), is first reconstructed through the hadronic decays $\bar D^0\to K^+\pi^-$, $K^+\pi^-\pi^0$, and
$K^+\pi^-\pi^-\pi^+$, which have large BFs and small background contamination. If a signal $D^0\to K^0_L\phi, K^0_L\eta, K^0_L\omega, {\rm or }K^0_L\eta^{\prime}$ decay can be reconstructed in the rest of the event, the event is then considered as a DT event. The BF of the signal decay is determined by
\begin{equation}
\label{eq:bf}
{\mathcal B}_{\rm sig}=N_{\mathrm{DT}}/(N_{\mathrm{ST}}^{\rm tot}\cdot \epsilon_{\rm sig}),
\end{equation}
where $N_{\rm ST}^{\rm tot}$ and $N_{\rm DT}$ are the yields of the total ST and DT candidates in data, respectively, and
$\epsilon_{\rm sig}=\Sigma_i[(\epsilon^i_{\rm DT}\cdot N^i_{\rm ST})/(\epsilon^i_{\rm
  ST}\cdot N_{\rm ST}^{\rm tot})]$ is the effective signal efficiency of finding the signal decay in the presence of the ST $\bar D^0$ meson,
where $\epsilon_{\rm ST}$ and $\epsilon_{\rm DT}$ are the detection efficiencies of the ST and DT candidates, respectively,
and the index $i$ runs over all ST modes.

In the work described in this paper, candidates for $K^\pm$, $\pi^\pm$, $\gamma$, and $\pi^0$ are selected by using the same selection criteria as in Ref.~\cite{bes3-etaX}.
The two-body ST mode $\bar D^0\to K^+\pi^-$ suffers from background contributions from cosmic rays and Bhabha scattering events. These background contributions are rejected by using the same requirements as in Ref.~\cite{deltakpi}. For the $\bar D^0\to K^+\pi^-\pi^-\pi^+$ ST mode, the $\bar D^0\to
K^0_SK^\pm\pi^\mp$ decays are rejected if the mass of any $\pi^+\pi^-$
pair falls within $(0.483,0.513)$ GeV/$c^2$.

Two kinematic variables, the energy difference $\Delta E\equiv E_{\bar D^0}-E_{\mathrm{beam}}$ and the beam-constrained mass $M_{\rm BC}\equiv\sqrt{E_{\mathrm{beam}}^{2}/c^4-|\vec{p}_{\bar D^0}|^{2}/c^2}$ are used to separate the ST $\bar D^0$ mesons from combinatorial backgrounds.
Here, $E_{\mathrm{beam}}$ is the beam energy and $E_{\bar D^0}$ and
$\vec{p}_{\bar D^0}$ denote the total energy and momentum of the ST $\bar D^0$
candidate in the $e^+e^-$ center-of-mass frame, respectively.
If there are multiple combinations in an event,
only the combination with the smallest $|\Delta E|$ is accepted.
To suppress combinatorial backgrounds, the $\Delta E$ of any ST candidate is required to be within $(-0.055,0.040)$ GeV for $\bar D^0\to K^+\pi^-\pi^0$ and within $(-0.025,0.025)$ GeV for $\bar D^0\to K^+\pi^-$ and $\bar D^0\to K^+\pi^-\pi^-\pi^+$.

\begin{figure*}[htbp]\centering
\includegraphics[width=1.0\linewidth]{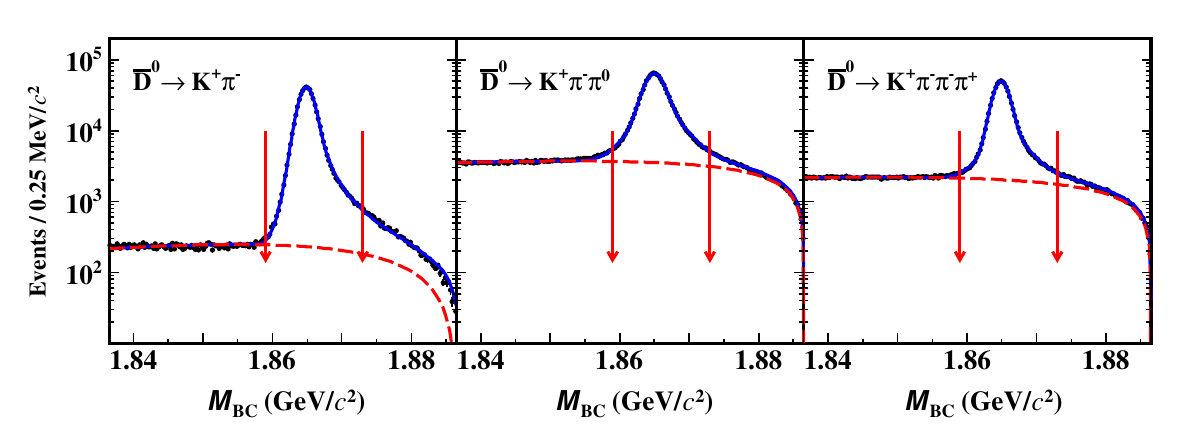}
\caption{ Fits to the $M_{\rm BC}$ distributions of the ST $\bar D^0$
  candidates.  Data are shown as dots (error bars are not visible at this scale).  The blue solid and
  red dashed curves are the total fit results and the fitted backgrounds,
  respectively.  Pairs of red arrows show the $M_{\rm BC}$ signal region.
}\label{fig:datafit_Massbc}
\end{figure*}

The $M_{\rm BC}$ distributions of the accepted ST $\bar D^0$ candidates are shown in Fig.~\ref{fig:datafit_Massbc}.
To extract the yield of ST $\bar D^0$ mesons for each ST mode, a binned maximum-likelihood fit is performed on the corresponding $M_{\rm BC}$ distribution.
The signal is modeled by the MC-simulated shape convolved with a double-Gaussian function to take into account the resolution difference between data and MC simulation. In the fits, the Gaussian means and widths are free parameters whose ranges are $(0.04,0.20)~{\rm MeV}/c^2$ and $(0.73,3.19)~{\rm MeV}/c^2$, respectively. The combinatorial background is described by the ARGUS function~\cite{argus}. The associated fit results are shown in Fig.~\ref{fig:datafit_Massbc}. The candidates with $M_{\rm BC}\in (1.859,1.873)$ GeV/$c^2$ are kept for further analyses.
Integrating the fitted signal shape in the aforementioned $M_{\rm BC}$ interval gives the yield of the ST $\bar D^0$ mesons for each ST mode. Summing over all ST modes, the total yield of ST $\bar D^0$ mesons is obtained to be $N^{\rm tot}_{\rm ST}=2266311\pm1842$.

The $D^0\to K^0_L X$ candidates are reconstructed with charged and photon candidates which have not been used in the ST side.
Candidates for $\phi$ and $\omega$ are reconstructed from $K^+K^-$ and $\pi^+\pi^-\pi^0$ combinations with $M_{K^+K^-}\in (1.005, 1.035)$ GeV/$c^2$ and $M_{\pi^+\pi^-\pi^0}\in (0.752, 0.812)$ GeV/$c^2$, respectively.
Candidates for $\eta$ are reconstructed from $\gamma\gamma$ pairs with
$M_{\gamma\gamma}\in (0.510,0.570)$\,GeV/$c^2$ or $\pi^+\pi^-\pi^0$ combinations with
$M_{\pi^+\pi^-\pi^0}\in (0.535,0.560)$\,GeV/$c^2$. Candidates for $\eta^{\prime}$ are reconstructed from $\pi^+\pi^-\eta~(\eta\to \gamma\gamma$) combinations with
$M_{\pi^+\pi^-\eta}\in (0.945,0.970)$\,GeV/$c^2$ or $\gamma\rho^0~(\rho^0\to\pi^+\pi^-)$ combinations with
$M_{\gamma\rho^0}\in (0.938,0.978)$\,GeV/$c^2$. To improve resolution of $\eta\to \gamma\gamma$, a mass constrained (1C) fit is performed, constraining the selected $\gamma\gamma$ pair invariant mass to the known $\eta$ mass~\cite{pdg2020}.
If there are multiple combinations of $\pi^0$ or $\eta_{\gamma\gamma}$, the one with the least $\chi^2_{\rm 1C}$ is kept for further analyses.
For $\eta^{\prime}\to \gamma\rho^0~(\rho^0\to\pi^+\pi^-)$, the $\pi^+\pi^-$ system is required to satisfy $M_{\pi^+\pi^-}\in (0.57, 0.97)$ GeV/$c^2$.
For $D^0\to K_L^0\eta^\prime_{\gamma\rho^0}$, the background events from $D^0\to K_L^0\pi^+\pi^-$ are suppressed by requiring that the recoil mass of the $\pi^+\pi^-$ pair from the signal combined with the ST particles is greater than 0.53 GeV/$c^2$. This requirement suppresses 90\% of background events from $D^0\to K_L^0\pi^+\pi^-$ at the cost of losing 0.5\% of the signal. If there are multiple $\gamma\rho^0$ combinations for $D^0\to K_L^0\eta^\prime_{\gamma\rho^0}$, the one with $M_{\gamma\rho^0}$ closest to the known $\eta^\prime$ mass~\cite{pdg2020} is kept for further analyses.
Throughout this paper, the subscripts of $\phi$, $\eta$, $\omega$, and $\eta^\prime$ denote the corresponding reconstruction modes.

To minimize the impact of a $K_L^0$ shower in electromagnetic calorimeter on an extra $\pi^0$ and $\eta$ vetoes,  the opening angle between any remaining photon and the missing momentum is required to be greater than 15$^\circ$. Events with extra charged tracks, $\pi^0$ or $\eta_{\gamma\gamma}$ are rejected to suppress background contributions from
$D^0\to K_S^0(\to \pi^+\pi^-)X$, $D^0\to K_S^0(\to\pi^0\pi^0) X$ and $D^0\to \eta_{\gamma\gamma}X$, respectively.

To separate signal events from background contributions, the variable ${\rm MM}^2 = E^2_{\rm miss}/c^4 - |\vec{p}_{\rm miss}|^2/c^2$ is defined, where $E_{\mathrm{miss}}$ and $\vec{p}_{\mathrm{miss}}$
are the missing energy and momentum of the DT event in the $e^+e^-$ center-of-mass frame, respectively.
They are calculated as $E_{\mathrm{miss}}\equiv E_{\mathrm{D^0}}-E_{X}$ and $\vec{p}_{\mathrm{miss}}\equiv\vec{p}_{D^0}-\vec{p}_{X}$, where $E_{D^0}$, $\vec{p}_{D^0}$, $E_{X}$ and $\vec{p}_{X}$ are the measured energy and momentum of the $D^0$ and $X$ candidates, respectively. The MM$^2$ resolution is improved by constraining the energy of $D^0$ to the beam energy and $\vec{p}_{D^0}\equiv-\hat{p}_{\bar D^0}\cdot$$\sqrt{E_{\mathrm{beam}}^{2}/c^4-m_{\bar D^0}^{2}}$, where
$\hat{p}_{\bar D^0}$ is the unit vector in the momentum direction of the ST $\bar D^0$ meson and $m_{\bar D^0}$ is the known $\bar D^0$ mass~\cite{pdg2020}.

\begin{figure*}[htbp] \centering
\includegraphics[width=1.0\linewidth]{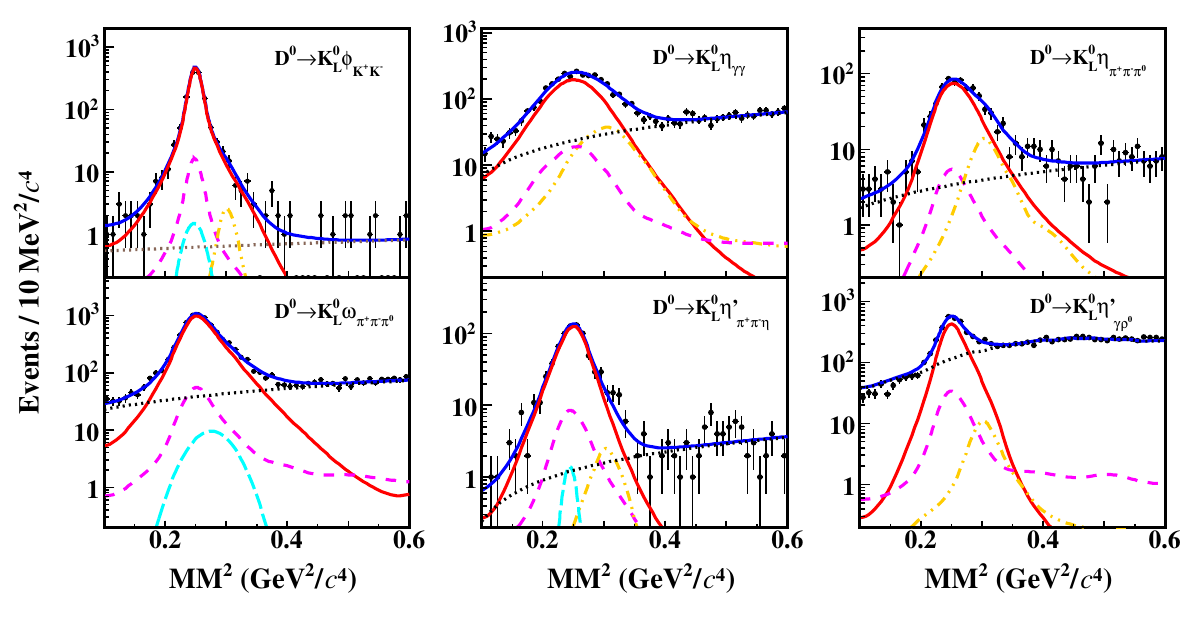}
\caption{
Fits to the MM$^2$ distributions of the accepted candidate events.
Data are shown as points with error bars.
Blue solid curves are the fit result, red solid curves are the signal,
pink dashed, yellow long-dashed-dotted,
light blue dashed-dotted, and black dotted curves denote BKGI, BKGII, BKGIII, and BKGIV, respectively.}
\label{fig:fit_Umistry1}
\end{figure*}

The signal yields ($N_{\rm sig}$) are extracted by fitting the MM$^2$ distributions of selected events.
Background events are divided into four categories.
The first  (BKGI) contains $D^0\to K_S^0(\to \pi^0\pi^0) X$ events.
The second (BKGII) contains $D^0\to\eta_{\gamma\gamma}\phi, \eta_{\gamma\gamma}\eta, \eta_{\gamma\gamma}\eta^{\prime}$ events.
The third (BKGIII) is from all the remaining peaking background channels.
The fourth (BKGIV) is from combinatorial background components.
In the fits,
the signal is modeled by the MC-simulated shape convolved with a double-Gaussian function. The means and widths of signal mode dependent Gaussian functions are in the intervals $(-0.58,1.97)~{\rm MeV^2}/c^4$ and $(0.16,3.70)~{\rm MeV^2}/c^4$, respectively.
The BKGI and BKGII are described by the corresponding MC-simulated shapes, and their sizes are fixed to the values estimated using the BFs from the PDG~\cite{pdg2020} and the corresponding misidentification rates.
The shape and size of BKGIII are fixed to those obtained from the inclusive MC sample.
BKGIV for $D^0\to K_L^0\eta^{\prime}_{\gamma\rho^0}$ is not smooth and is modeled by the MC-simulated shape; it is modeled by a linear function for the other signal decays.
Figure~\ref{fig:fit_Umistry1} shows the results of the fits to the MM$^2$ distributions of the accepted candidates in data.

There are combinatorial backgrounds in the $\phi \to K^+K^-$, $\eta \to \pi^+\pi^-\pi^0$, $\omega \to \pi^+\pi^-\pi^0$, and  $\eta^\prime \to \pi^+\pi^-\eta$ signal regions, which can form a peak in the $\rm MM^2$ distributions. This kind of background is estimated by the corresponding sideband regions, defined as $M_{K^+K^-}\in(0.985, 1.000)\bigcup(1.045, 1.060)$ GeV/$c^2$, $M_{\pi^+\pi^-\pi^0}\in(0.495, 0.515)\bigcup(0.580, 0.600)$ GeV/$c^2$, $M_{\pi^+\pi^-\pi^0}\in(0.712, 0.732)\bigcup(0.832, 0.852)$ GeV/$c^2$, $M_{\pi^+\pi^-\eta}\in(0.913, 0.938)\bigcup(0.978, 1.003)$ GeV/$c^2$. Their yields ($N_{\rm sid}$) are obtained from similar fits to individual MM$^2$ distributions.
Table~\ref{tab:bf} summarizes the fitted yields of $N_{\rm sig}$ and $N_{\rm sid}$, the normalization factors of background events in the signal and sideband regions ($S_{\rm co}$),
the net signal yield ($N_{\rm net}=N_{\rm sig}-N_{\rm sid}\cdot S_{\rm co}$), the signal efficiencies ($\epsilon_{\rm sig}$), and the obtained BFs ($\mathcal{B}_{\rm sig}$).

At $\sqrt{s}=3.773$ GeV, the $D^0\bar D^0$ pairs are produced coherently.
The measurements of BFs with the DT method are affected by the quantum correlation (QC) effect.
Following Ref.~\cite{D-phix}, this effect is considered as a tag-mode-dependent correction factor, $f_{\rm QC}^i=\frac{1}{1-C_f^i\cdot(2F_+^{\rm sig}-1)}$,
where $C_f^i$ is the strong-phase factor calculated as $C_f^i=\frac{2r_iR_i{\rm cos}\delta_f^i}{1+r_i^2}$,
$R_i$ is the coherence factor,
$\delta_f^i$ is the strong-phase difference between the CF and DCS amplitudes,
$r_f^i$ is defined as $r^i_fe^{-i\delta^i_f}\equiv\frac{\langle f|\bar{D^0}\rangle}{\langle f|D^0\rangle}$,
for the tag mode $i$; and $F_+^{\rm sig}$ is the $CP+$ fraction for the signal decay and it equals to 1 for all the studied decays.
With necessary parameters quoted from Refs.~\cite{fqc1,fqc2},
the $f_{\rm QC}^i$ factors are determined to be $0.898\pm0.007$, $0.935\pm0.007$, and $0.972\pm0.019$ for $D\to K^-\pi^+$, $D\to K^-\pi^+\pi^0$, and $D\to K^-\pi^+\pi^+\pi^-$, respectively.
All signal decay final states studied are $CP+$ eigenstates.
The averaged QC correction factor,  which has been weighted by the ST yields in data, is determined to be $f_{\rm QC}=0.937\pm0.007$.
Multiplying the directly-measured BFs by this factor yields the reported BFs.
After this correction, the residual uncertainty of $f_{\rm QC}$ will be assigned as a systematic uncertainty.
For $D^0\to K_L^0\eta$ and $D^0\to K_L^0\eta^\prime$, the BFs measured by two different $\eta$ or $\eta^{\prime}$ decay modes have been weighted by
the combined statistical and independent uncertainties and the obtained results are shown in Table~\ref{tab:bf}.

\begin{table*}[htpb]
  \centering
  \caption{
  The quantities used for BF determinations and the obtained BFs.
  The signal efficiencies include the BFs for all possible sub-decays and necessary correction factors mentioned later.
  The listed BFs have been corrected by the QC factor $f_{\rm QC}$ and
  $\bar {\mathcal B} $ denotes the weighted average BFs for $D^0\to K_L^0\eta$ and $D^0\to K_L^0\eta^\prime$. The first and second uncertainties for $\mathcal B $ are statistical and systematic, respectively; uncertainties for other variables are statistical only. 
  }
  \scalebox{0.96}{
  \begin{tabular}{l|c|c|c|c|c|c|c}
  \hline
  \hline
  \multicolumn{1}{l|} {Decay} & $N_{\rm sig}$ & $N_{\rm sid}$ & $S_{\rm co}$ & $N_{\rm net}$ & $\epsilon_{\rm sig} (\%)$ & $\mathcal{B}_{\rm sig}~(\%)$ & $\bar {\mathcal B}_{\rm sig}~(\%)$ \\
  \hline
  $D^0\to K_L^0\phi_{K^+K^-}$ & $1271\pm39$ & $276\pm19$ & $1.33\pm0.17$ & $904\pm46$ & $9.02\pm0.10$  & $0.414\pm0.021\pm0.010$& ... \\ \hline
  $D^0\to K_L^0\eta_{\gamma\gamma}$ & $2132\pm71$ & ... & ... & $2132\pm71$ & $20.46\pm0.15$  & $0.431\pm0.014\pm0.013$&  \multirow{2}{*}{$0.433\pm 0.012\pm 0.010$} \\
  $D^0\to K_L^0\eta_{\pi^+\pi^-\pi^0}$ & $565\pm29$ & $36\pm10$ & $0.61\pm0.10$ & $543\pm30$ & $5.11\pm0.07$  & $0.439\pm0.024\pm0.015$& \\ \hline
  $D^0\to K_L^0\omega_{\pi^+\pi^-\pi^0}$ & $6692\pm100$ & $368\pm39$ & $1.58\pm0.07$ & $6110\pm118$ & $21.70\pm0.18$  & $1.164\pm0.022\pm0.028$& ...\\ \hline
  $D^0\to K_L^0\eta_{\pi^+\pi^-\eta}^{\prime}$ & $688\pm29$ & $8\pm6$ & $0.47\pm0.08$ & $684\pm29$ & $3.30\pm0.10$ & $0.857\pm0.037\pm0.022$&\multirow{2}{*}{$0.809\pm 0.020\pm 0.016$} \\
  $D^0\to K_L^0\eta_{\rho^0\gamma}^{\prime}$ & $2002\pm61$ & ... & ... & $2002\pm61$ & $10.55\pm0.15$  & $0.785\pm0.024\pm0.023$& \\
  \hline
  \hline
  \end{tabular}}
  \label{tab:bf}
\end{table*}

In the measurements of the BFs for $D^0\to K_L^0 X$ using the DT method, the systematic uncertainties associated with the ST selection are canceled. The major sources of systematic uncertainties related to the measured BFs are described below.

The uncertainty in the total yield of ST $\bar D^0$ mesons has been studied in Ref.~\cite{epjc76} and is evaluated as 0.5\%.
The tracking and particle identification (PID) efficiencies of charged kaons and pions are studied by analyzing DT hadronic $D\bar D$ events~\cite{D-PP}. The data/MC differences in various momentum intervals are re-weighted by the corresponding momentum distributions of the signal decays. We correct the MC efficiencies to data by signal mode dependent factors of $(0.2-5.5)\%$, where the larger difference between data and MC simulation comes from the tracking efficiencies for low momentum $K$ in $D^0\to K_L^0\phi_{K^+K^-}$ decay. The residual systematic uncertainties are $(0.2-0.6)\%$ for tracking and PID efficiencies per $K^\pm$ or $\pi^\pm$.

The systematic uncertainty due to the photon detection in $D^0\to K_L^0\eta^{\prime}_{\rho^0\gamma}$ decay is 1.0\% per photon, as estimated from a $J/\psi\to\rho^0\pi^0$ control sample~\cite{gam_sys}

The systematic uncertainty of $\pi^0$ reconstruction has been studied by using the DT events of $\bar{D}^0\to K^+\pi^-$, $K^+\pi^-\pi^-\pi^+$ vs.~$D^0\to K^-\pi^+\pi^0$ and $K_S^0\pi^0$~\cite{epjc76}.  After correcting the MC efficiencies to agree with data using the momentum-weighted difference $(0.5-0.6)\%$, the residual systematic uncertainty, 0.7\%, is assigned as the systematic uncertainty in $\pi^0$ reconstruction. Due to the limited size of the $\eta$ sample, the uncertainties of the $\eta$ reconstruction in $D^0\to K_L^0\eta_{\gamma\gamma}$ and $D^0\to K_L^0\eta^{\prime}_{\pi^+\pi^-\eta}$ decays are assigned to be 0.5\% and 0.7\% by referring to the $\pi^0$ reconstruction.

The systematic uncertainties due to the mass windows of $\phi_{K^+K^-}$, $\eta_{\pi^+\pi^-\pi^0}$, $\omega_{\pi^+\pi^-\pi^0}$,  $\eta^{\prime}_{\pi^+\pi^-\eta}$, and $\eta^{\prime}_{\rho^0\gamma}$ candidates are studied using control samples of $D^0\to K_S^0 \phi_{K^+K^-}$, $ K_S^0 \eta_{\pi^+\pi^-\pi^0}$, $ K_S^0 \omega_{\pi^+\pi^-\pi^0}$,  $ K_S^0 \eta^{\prime}_{\pi^+\pi^-\eta}$, and $ K_S^0 \eta^{\prime}_{\rho^0\gamma}$, respectively. The relative differences of $(0.2-0.5)\%$ in the acceptance efficiencies between data and MC simulation are taken as individual systematic uncertainties.

The systematic uncertainty due to requiring no extra charged track, $\pi^0$ and $\eta$ is studied using the control sample of $D^0\to K_S^0\pi^0$. The relative difference in efficiencies between data and MC simulation, 0.8\%, is assigned as the systematic uncertainty.

The systematic uncertainties arising from the MM$^2$ fits are evaluated by varying the signal shape, the background shape, and the size of peaking backgrounds within their uncertainties.
The relative changes of various re-measured BFs are added in quadrature and these totals, $(0.9-2.4)\%$, are taken as the corresponding systematic uncertainties.

The BFs with alternative sideband regions of $M_{K^+K^-}$ and $M_{\pi^+\pi^-\pi^0}$ ($\pm10$ MeV$/c^2$ or $\pm15$ MeV$/c^2$)
as well as the requirement of $M_{\pi^+\pi^-}^{\rm recoil}$ (enlarged by 20 MeV/$c^2$) are examined.
Accounting for correlations, changes in the re-measured BFs are negligible.

To estimate the systematic uncertainties due to the requirement of the opening angle between any remaining shower and the missing momentum of $\bar{D}^0X$, the BFs are measured using different angle requirements. The maximum deviation of the BFs from $(0.9-1.6)\%$ is taken as the systematic uncertainty.

The uncertainties of the quoted BFs~\cite{pdg2020} of $\phi\to K^+K^-$, $\eta\to \gamma\gamma$, $\eta\to \pi^+\pi^-\pi^0$, $\omega\to \pi^+\pi^-\pi^0$,
$\eta^\prime\to \pi^+\pi^-\eta$, and $\eta^\prime\to \rho^{0}\gamma$ are assigned as individual systematic uncertainties.

The QC effect on the measured BFs has been corrected by the factor $f_{\rm QC}$
aforementioned and the residual error of $f_{\rm QC}$ is assigned as the systematic uncertainty, which is 0.7\%.

The uncertainties arising from the finite sizes of the signal MC samples are $(0.3-0.6)\%$.
Systematic uncertainties from other selection criteria are found to be negligible.

For each signal decay, the total systematic uncertainty is obtained by summing individual contributions in quadrature and is shown in Table~\ref{tab:system_total}.

\begin{table*}[htpb]
  \centering
  \caption{Systematic uncertainties (\%) in the measurements of the BFs.}
  \scalebox{1.0}{
  \begin{tabular}{l|c|c|c|c|c|c}
  \hline
  \hline
  Source & $K_L^0\phi_{K^+K^-}$ & $K_L^0\eta_{\gamma\gamma}$ & $K_L^0\eta_{\pi^+\pi^-\pi^0}$ & $K_L^0\omega_{\pi^+\pi^-\pi^0}$ & $K_L^0\eta^{\prime}_{\pi^+\pi^-\eta}$ & $K_L^0\eta^{\prime}_{\rho^0\gamma}$ \\
  \hline
  ST yield $N_{\rm tag}$       & 0.5 & 0.5 & 0.5 & 0.5 & 0.5 & 0.5 \\
  $K^{\pm}/\pi^{\pm}$ tracking  & 0.4 & -   & 0.3 & 0.2 & 0.6 & 0.2 \\
  $K^{\pm}/\pi^{\pm}$ PID       & 0.3 & -   & 0.2 & 0.2 & 0.2 & 0.2 \\
  $\gamma$ reconstruction       & - & - & - & - & - & 1.0 \\
  $\pi^0/\eta$ reconstruction   & - & 0.5 & 0.7 & 0.7 & 0.7 & - \\
  Mass window requirement       & 0.2 & - & 0.2 & 0.1 & 0.5 & 0.1 \\
  $N_{\rm charged/\pi^0/\eta}^{\rm extra}$ & 0.8 & 0.8 & 0.8 & 0.8 & 0.8 & 0.8 \\
  MM$^2$ fit                 & 0.9 & 2.3 & 2.4 & 1.1 & 1.4 & 1.2\\
  Opening angle                    & 1.4 & 1.4 & 1.4 & 1.3 & 0.9 & 1.6 \\
  Quoted BFs                    & 1.0 & 0.5 & 1.2 & 0.7 & 1.2 & 1.3 \\
  MC statistics                 & 0.6 & 0.3 & 0.5 & 0.5 & 0.6 & 0.4 \\
  Strong phase                  & 0.7 & 0.7 & 0.7 & 0.7 & 0.7 & 0.7 \\
  \hline
  Total                         & 2.4 & 3.0 & 3.4 & 2.4 & 2.6 & 2.9 \\
  \hline
  \hline
  \end{tabular}}
  \label{tab:system_total}
\end{table*}

\begin{table}[htp]
\centering
\caption{\small
$CP$-conjugate BFs (${\mathcal B}^+_{\rm sig}$ and ${\mathcal B}^-_{{\rm sig}}$)
and their asymmetries (${{\mathcal A}_{CP}^{\rm sig}}$).
The first and second uncertainties are statistical and systematic, respectively.
}\label{tab:CP}
\scalebox{0.95}{
\begin{tabular}{c|c|c|c}
  \hline\hline
{Decay} & ${\mathcal B}^+_{\rm sig}$ (\%)  & ${\mathcal B}^-_{{\rm sig}}$ (\%)  & ${\mathcal A}_{CP}^{\rm sig}$ (\%) \\  \hline
$D^0\to K_L^0\phi$          &$0.428\pm0.029$&$0.405\pm0.034$&$ 2.7\pm5.4\pm0.7$\\
$D^0\to K_L^0\eta$          &$0.445\pm0.018$&$0.421\pm0.017$&$ 2.8\pm2.9\pm0.4$\\
$D^0\to K_L^0\omega$        &$1.200\pm0.030$&$1.121\pm0.031$&$ 3.4\pm1.9\pm0.6$\\
$D^0\to K_L^0\eta^{\prime}$ &$0.789\pm0.028$&$0.826\pm0.028$&$-2.2\pm2.5\pm0.4$\\
\hline\hline
\end{tabular}
}
\end{table}

\begin{table*}[htpb]
  \centering
  \caption{Comparison of measured BFs and $K_S^0$-$K_L^0$ asymmetries with theoretical calculations of Ref.~\cite{lanzhou_theory}. $\mathcal{B}_{\rm exp}$ ($\mathcal{B}_{\rm FAT}$) and $\mathcal{R}(D^0)_{\rm exp}$ ($\mathcal{R}(D^0)_{\rm FAT}$) are the BFs and $K_S^0$-$K_L^0$ asymmetries of the experimental measurements (theoretical calculations).}

  \begin{tabular}{l||c|c|c||c|c|c}
  \hline
  \hline
\multicolumn{1}{c||} {Decay} & $\mathcal{B}_{\rm exp}$ (\%) & $\mathcal{B}_{\rm FAT}$  (\%) & Difference & $\mathcal{R}(D^0)_{\rm exp}$ & $\mathcal{B}(D^0)_{\rm FAT}$ & Difference \\
  \hline
  $D^0\to K_L^0\phi$   & $0.414\pm0.021\pm0.010$ & $0.33\pm0.03$ & $2.2\sigma$ & $-0.001\pm0.047$  & \multirow{4}{*}{$0.113\pm0.001$} & $2.4\sigma$\\
  $D^0\to K_L^0\eta$   & $0.433\pm0.012\pm0.010$ & $0.40\pm0.07$ & $0.5\sigma$ & $0.080\pm0.022$   &  & $1.5\sigma$\\
  $D^0\to K_L^0\omega$ & $1.164\pm0.022\pm0.028$ & $0.95\pm0.15$ & $1.4\sigma$ & $-0.024\pm0.031$  &  & $4.4\sigma$\\
  $D^0\to K_L^0\eta'$  & $0.809\pm0.020\pm0.016$ & $0.77\pm0.07$ & $0.5\sigma$ & $0.080\pm0.023$   &  & $1.6\sigma$\\
  \hline
  \hline
  \end{tabular}
  \label{tab:final}
\end{table*}

The BFs of $D$ and $\bar D$ decays, ${\mathcal B}^+_{\rm sig}$ and ${\mathcal B}^-_{{\rm sig}}$, are also measured separately. Their asymmetry is determined by
${{\mathcal A}_{CP}^{\rm sig}}=\frac{{\mathcal B}^+_{\rm sig}-{\mathcal B}^-_{{\rm sig}}}{{\mathcal B}^+_{\rm sig}+{\mathcal B}^-_{{\rm sig}}}$.
The obtained BFs and asymmetries are summarized in Table~\ref{tab:CP}. No significant $CP$ violation is observed.
Several systematic uncertainties cancel in the asymmetry, such as
the tracking and PID of $\pi^+\pi^-$, $K^+K^-$ pairs, $\pi^0, \eta$ reconstruction, quoted BFs, $K^0_S, \omega, \eta^{(\prime)}, \phi$ sideband choices,
and the strong phase between $D^0$ and $\bar D^0$ decays. The other systematic uncertainties are estimated separately as above.

In summary, by analyzing 2.93 fb$^{-1}$ of $e^+e^-$ annihilation data taken at $\sqrt{s}=3.773$ GeV with the BESIII detector, we have performed the first measurements of the absolute BFs of $D^0\to K_L^0\phi$, $D^0\to K_L^0\eta$, $D^0\to K_L^0\omega$, and $D^0\to K_L^0\eta^{\prime}$.
Combining the BFs measured in this work with the known values for $\mathcal{B}(D^0\to K_S^0 X)$~\cite{pdg2020}, the asymmetries of $\mathcal{B}(D^0\to K_S^0 X)$ and $\mathcal{B}(D^0\to K_L^0 X)$ are determined. Table~\ref{tab:final} shows the comparison of the measured BFs and $K_S^0$-$K_L^0$ asymmetries with the theoretical calculations of Ref.~\cite{lanzhou_theory}. Clear asymmetries are found in $D^0\to K_L^0\eta, K_L^0\eta^{\prime}$, but none is found in the other two modes.
Our results ${\mathcal R}(D^0,\eta)=0.080\pm0.022$ and ${\mathcal R}(D^0,\eta^{\prime})=0.080\pm0.023$ are consistent with ${\mathcal R}(D^0,\pi^0)=0.108\pm0.035$ measured by CLEO~\cite{cleo_exp}
and imply that the $K_S^0$-$K_L^0$ asymmetry in $D^0\to K^0_{S/L}\eta, K^0_{S/L}\eta^{\prime}$ modes is approximately $2\tan^2\theta_C$ as expected based on SU(3) symmetry~\cite{asymmetry_theory1,Rosner:2006bw,asymmetry_theory2,asymmetry_theory3,lanzhou_theory}.
Comparing with the ${\mathcal R}(D^+,\pi^+)$~\cite{cleo_exp} and ${\mathcal R}(D^+_s,K^+)$~\cite{bes_Ds}, a significantly larger asymmetry for ${\mathcal R}(D^0,X)$ is expected due to a smaller strong phase difference between the DCS and CF amplitudes.
However, the obtained $K_S^0$-$K_L^0$ asymmetries in $D^0\to K_{S,L}^0\phi$ and $K_{S,L}^0\omega$ decays disagree with the predicted value~\cite{lanzhou_theory} by $2.4\sigma$ and $4.4\sigma$, respectively.
The main possible reason of this tension is that the $E_P=E_V$ assumption in Ref.~\cite{lanzhou_theory} is not satisfied.
In addition, the asymmetries of the $CP$-conjugate BFs for these $D$ decays are determined
and no significant $CP$ violation is found.
These results offer crucial information to more reliably calculate the BFs of the $D\to PP$ and $D\to VP$ decays in theories and
will aid investigations of quark SU(3)-flavor symmetry breaking as well as $CP$ violation in the hadronic decays of charmed mesons~\cite{bes3-white-paper, lihb}.
Our $K_S^0$-$K_L^0$ asymmetries offer the first opportunity to access individual amplitudes of DCS processes involving $K^0$
which can be only measured with quantum correlated $e^+e^-\to D\bar{D}$ production near the threshold, thereby further restricting the $D^0-\bar D^0$ mixing effect in charm decays.

The BESIII collaboration thanks the staff of BEPCII and the IHEP computing center for their strong support. This work is supported in part by the National Key R\&D Program of China under Grants Nos. 2020YFA0406400 and No. 2020YFA0406300; the National Natural Science Foundation of China (NSFC) under Grants No. 11775230, No. 12035009, No. 11875170, No. 11475090, No. 11625523, No. 11635010, No. 11735014, No. 11822506, No. 11835012, No. 11935015, No. 11935016, No. 11935018, No. 11961141012, No. 12022510, No. 12025502, No. 12035013, No. 12061131003, No. 12192260, No. 12192261, No. 12192262, No. 12192263, No. 12192264, and No. 12192265; the Chinese Academy of Sciences (CAS) Large-Scale Scientific Facility Program; Joint Large-Scale Scientific Facility Funds of the NSFC and CAS under Grants No. U1732263 and No. U1832207; CAS Key Research Program of Frontier Sciences under Grant No. QYZDJ-SSW-SLH040; 100 Talents Program of CAS; INPAC and Shanghai Key Laboratory for Particle Physics and Cosmology; ERC under Grant No. 758462; European Union Horizon 2020 research and innovation programme under Marie Sklodowska-Curie Grant Agreement No. 894790; German Research Foundation DFG under Grant No. 443159800, Collaborative Research Center CRC 1044, FOR 2359, GRK 214; Istituto Nazionale di Fisica Nucleare, Italy; Ministry of Development of Turkey under Grant No. DPT2006K-120470; National Science and Technology fund; Olle Engkvist Foundation under Grant No. 200-0605; STFC (United Kingdom); The Knut and Alice Wallenberg Foundation (Sweden) under Grant No. 2016.0157; The Royal Society, UK under Grants No. DH140054 and No. DH160214; The Swedish Research Council; and U. S. Department of Energy under Awards No. DE-FG02-05ER41374 and DE-SC-0012069.

\clearpage

%
%
%


\begin{thebibliography}{99}

\bibitem{pdg2020}
P. A. Zyla {\it et al.} (Particle Data Group),
\href{http://pdglivetest.lbl.gov/Viewer.action}{Prog. Theor. Exp. Phys. {\bf 2020}, 083C01 (2020).}

\bibitem{asymmetry_theory1}
I. I. Bigi and H. Yamamoto,
\href{https://doi.org/10.1016/0370-2693(95)00285-S}{Phys. Lett. B {\bf 349}, 363 (1995).}

\bibitem{Rosner:2006bw}
J. L. Rosner,
\href{doi:10.1103/PhysRevD.74.057502}{Phys. Rev. D \textbf{74}, 057502 (2006).}

\bibitem{asymmetry_theory2}
B. Bhattacharya and J. L. Rosner,
\href{https://journals.aps.org/prd/abstract/10.1103/PhysRevD.81.014026}{Phys. Rev. D {\bf 81}, 014026 (2010).}

\bibitem{asymmetry_theory3}
D. N. Gao,
\href{https://journals.aps.org/prd/abstract/10.1103/PhysRevD.91.014019}{Phys. Rev. D {\bf 91}, 014019 (2015).}

\bibitem{lanzhou_theory}
D. Wang, F. S. Yu, P. F. Guo, and H. Y. Jiang,
\href{https://journals.aps.org/prd/abstract/10.1103/PhysRevD.95.073007}{Phys. Rev. D {\bf 95}, 073007 (2017).}

\bibitem{Yu:2017oky}
D. Wang, F. S. Yu, and H. N. Li,
\href{doi:10.1103/PhysRevLett.119.181802}{Phys. Rev. Lett. \textbf{119}, 181802 (2017).}


\bibitem{mixing_parameter} Y. Amhis {\it et al.} (HFLAV),
\href{https://link.springer.com/article/10.1140\%2Fepjc\%2Fs10052-017-5058-4}{Eur. Phys. J. C {\bf 77}, 895 (2017).}

\bibitem{asy_explain}
D. N. Gao,
\href{https://reader.elsevier.com/reader/sd/pii/S0370269306015322?token=1D278FA804400B929D02543F871BA7425B858260E15BF3CB1D50C5AD196BBF7090E5A1AF1E595DC7348D4BBFB155B8F5&originRegion=us-east-1&originCreation=20210726065032}{Phys. Lett. B {\bf 645}, 59 (2007).}

\bibitem{cleo_exp}
Q. He {\it et al.} (CLEO Collaboration),
\href{https://journals.aps.org/prl/abstract/10.1103/PhysRevLett.100.091801}{Phys. Rev. Lett. {\bf 100}, 091801 (2008).}

\bibitem{Kingsley}
R. L. Kingsley, S. B. Treiman, F. Wilczek and A. Zee,
\href{https://journals.aps.org/prd/abstract/10.1103/PhysRevD.11.1919}{Phys. Rev. D {\bf 11}, 1919 (1975).}

\bibitem{Gronau}
M. Gronau,
\href{https://doi.org/10.1016/j.physletb.2014.01.035}{Phys. Lett. B {\bf 730}, 221 (2014);}
\href{https://doi.org/10.1016/j.physletb.2014.06.055}{Erratum: Phys. Lett. B {\bf 735}, 282 (2014).}

\bibitem{theory_a}
H. J. Lipkin,
\href{https://journals.aps.org/prl/abstract/10.1103/PhysRevLett.46.1307}{Phys. Rev. Lett. {\bf 46}, 1307 (1981).}

\bibitem{theory_1}
H. Y. Cheng and C. W. Chiang,
\href{https://journals.aps.org/prd/abstract/10.1103/PhysRevD.81.074021}{ Phys. Rev. D {\bf 81}, 074021 (2010).}

\bibitem{theory_2}
Q. Qin, H. N. Li, C. D. L\"u, and F. S. Yu,
\href{https://journals.aps.org/prd/abstract/10.1103/PhysRevD.89.054006}{Phys. Rev. D {\bf 89}, 054006 (2014).}

\bibitem{theory_3}
H. Y. Cheng, C. W. Chiang, and A. L. Kuo,
\href{https://journals.aps.org/prd/abstract/10.1103/PhysRevD.93.114010}{Phys. Rev. D {\bf 93}, 114010 (2016).}

\bibitem{theory_4}
W. Kwong and S. P. Rosen,
\href{https://doi.org/10.1016/0370-2693(93)91843-C}{Phys. Lett. B {\bf 298}, 413 (1993).}

\bibitem{theory_5}
Y. Grossman and D. J. Robinson,
\href{https://doi.org/10.1007/JHEP04(2013)067}{J. High Energy Phys. {\bf 04}, 067 (2013).}

\bibitem{ref5}
H. N. Li, C. D. L\"u, and F. S. Yu,
\href{https://journals.aps.org/prd/abstract/10.1103/PhysRevD.86.036012}{Phys. Rev. D {\bf 86}, 036012 (2012).}

\bibitem{qqin}
Q. Qin, C. Wang, D. Wang, S. H. Zhou,
\href{https://arxiv.org/abs/2111.14472}{arXiv:2111.14472[hep-ph].}

\bibitem{zzxing}
Z. Z. Xing,
\href{https://doi.org/10.1142/S0217732319502389}{Mod. Phys. Lett. A {\bf 34}, 1950238 (2019).}

\bibitem{gamma_angle1}
M. Ablikim {\it et al.} (BESIII Collaboration),
\href{https://journals.aps.org/prl/abstract/10.1103/PhysRevLett.124.241802}{Phys. Rev. Lett. {\bf 124}, 241802 (2020).}

\bibitem{gamma_angle2}
M. Ablikim {\it et al.} (BESIII Collaboration),
\href{https://doi.org/10.1007/JHEP05(2021)164}{J. High Energy Phys. {\bf 05}, 164 (2021).}

\bibitem{ref1}
I. I. Bigi, A. Paul, and S. Recksiegel,
\href{https://doi.org/10.1007/JHEP06(2011)089}{J. High Energy Phys. {\bf 06}, 089 (2011).}

\bibitem{ref2}
G. Isidori, J. F. Kamenik, Z. Ligeti, and G. Perez,
\href{https://doi.org/10.1016/j.physletb.2012.03.046}{Phys. Lett. B {\bf 711}, 46 (2012).}

\bibitem{ref3}
J. Brod, A. L. Kagan, and J. Zupan,
\href{https://journals.aps.org/prd/abstract/10.1103/PhysRevD.86.014023}{Phys. Rev. D {\bf 86}, 014023 (2012).}

\bibitem{ref4}
H. Y. Cheng and C. W. Chiang,
\href{https://journals.aps.org/prd/abstract/10.1103/PhysRevD.86.014014}{Phys. Rev. D {\bf 86}, 014014 (2012).}

\bibitem{ref6}
H. Y. Cheng and C. W. Chiang,
\href{https://journals.aps.org/prd/abstract/10.1103/PhysRevD.100.093002}{Phys. Rev. D {\bf 100}, 093002 (2019).}

\bibitem{ref7}
H. N. Li, C. D. L\"u, and F. S. Yu,
\href{https://arxiv.org/abs/1903.10638}{arXiv:1903.10638[hep-ph].}

\bibitem{lhcb_D_CP}
R. Aaij {\it et al}. (LHCb Collaboration),
\href{https://journals.aps.org/prl/abstract/10.1103/PhysRevLett.122.211803}{Phys. Rev. Lett. {\bf 122}, 211803 (2019).}

\bibitem{lum_bes3}
M. Ablikim {\it et al.} (BESIII Collaboration),
\href{https://iopscience.iop.org/article/10.1088/1674-1137/37/12/123001}{Chin. Phys. C {\bf 37}, 123001 (2013);}
\href{https://doi.org/10.1016/j.physletb.2015.11.043}{Phys. Lett. B {\bf 753}, 629 (2016).}

\bibitem{BESCol} M. Ablikim {\it et al.} (BESIII Collaboration),
\href{https://doi.org/10.1016/j.nima.2009.12.050}{Nucl. Instr. Method A {\bf 614}, 345 (2010).}

\bibitem{geant4} S. Agostinelli {\it et al.} (GEANT4 Collaboration),
\href{https://doi.org/10.1016/S0168-9002(03)01368-8}{Nucl. Instr. Method A {\bf 506}, 250 (2003).}

\bibitem{kkmc} S. Jadach, B. F. L. Ward, and Z. Was,
\href{https://journals.aps.org/prd/abstract/10.1103/PhysRevD.63.113009} {Phys. Rev. D {\bf 63}, 113009 (2001);}
\href{https://doi.org/10.1016/S0010-4655(00)00048-5}{Comput. Phys. Commun.  {\bf 130}, 260 (2000).}

\bibitem{evtgen}
D.~J.~Lange,
\href{https://doi.org/10.1016/S0168-9002(01)00089-4} {Nucl. Instrum. Meth. A {\bf 462}, 152 (2001);}
R.~G.~Ping,
\href{https://doi.org/10.1088/1674-1137/32/8/001}{Chin. Phys. C {\bf 32}, 599 (2008).}

\bibitem{lundcharm}J. C. Chen, G. S. Huang, X. R. Qi, D. H. Zhang, and Y. S. Zhu,
\href{https://journals.aps.org/prd/abstract/10.1103/PhysRevD.62.034003}{Phys. Rev. D {\bf 62}, 034003 (2000);} 
R. L. Yang, R. G. Ping and H. Chen, \href{https://doi.org/ 10.1088/0256-307X/31/6/061301} {Chin. Phys. Lett.{\bf 31}, 061301 (2014).}

\bibitem{photos}
E.~Richter-Was,
\href{https://doi.org/10.1016/0370-2693(93)90062-M}{Phys. Lett. B {\bf 303}, 163 (1993).}

\bibitem{DT_method}
R. M. Baltrusaitis {\it et al.} (MARKIII Collaboration),
\href{https://journals.aps.org/prl/abstract/10.1103/PhysRevLett.56.2140}{Phys. Rev. Lett. {\bf 56}, 2140 (1986).}

\bibitem{bes3-etaX}
M. Ablikim {\it et al.} (BESIII Collaboration),
\href{https://journals.aps.org/prl/pdf/10.1103/PhysRevLett.124.241803}{Phys. Rev. Lett. {\bf 124}, 241803 (2020).}

\bibitem{deltakpi}
M. Ablikim {\it et al.} (BESIII Collaboration),
\href{https://doi.org/10.1016/j.physletb.2014.05.071}{Phys. Lett. B {\bf 734}, 227 (2014).}

\bibitem{argus} H. Albrecht {\it et al.} (ARGUS Collaboration),
\href{https://doi.org/10.1016/0370-2693(90)91293-K}{Phys. Lett. B {\bf 241}, 278 (1990).}

\bibitem{D-phix}
M. Ablikim {\it et al.} (BESIII Collaboration),
\href{https://journals.aps.org/prd/pdf/10.1103/PhysRevD.100.072006}{Phys. Rev. D {\bf 100}, 072006 (2019).}

\bibitem{fqc1}
T. Evans, S. T. Harnew, J. Libby, S. Malde, J. Rademacker, and G. Wilkinson,
\href{https://www.sciencedirect.com/science/article/pii/S0370269316301101?via\%3Dihub}{Phys. Lett. B {\bf 757}, 520 (2016).}

\bibitem{fqc2}
Heavy Flavor Averaging Group (HFLAV),
\href{https://hflav.web.cern.ch/content/charm-physics}{https://hflav.web.cern.ch/content/charm-physics}


\bibitem{epjc76} M. Ablikim {\it et al.} (BESIII Collaboration),
\href{https://doi.org/10.1140/epjc/s10052-016-4198-2} {Eur. Phys. J. C {\bf 76}, 369 (2016).}


\bibitem{D-PP}
M. Ablikim {\it et al.} (BESIII Collaboration),
\href{https://journals.aps.org/prd/pdf/10.1103/PhysRevD.97.072004}{Phys. Rev. D {\bf 97}, 072004 (2018).}


\bibitem{gam_sys}
M. Ablikim {\it et al.} (BESIII Collaboration),
\href{https://journals.aps.org/prd/abstract/10.1103/PhysRevD.86.052011}{Phys. Rev. D {\bf 86}, 052011 (2012).}



\bibitem{bes_Ds}
M. Ablikim {\it et al.} (BESIII Collaboration),
\href{https://journals.aps.org/prd/abstract/10.1103/PhysRevD.99.112005}{Phys. Rev. D {\bf 99}, 112005 (2019).}

\bibitem{bes3-white-paper} M. Ablikim {\it et al.} (BESIII Collaboration),
\href{http://cpc.ihep.ac.cn/article/doi/10.1088/1674-1137/44/4/040001} {Chin. Phys. C {\bf 44}, 040001 (2020).}

\bibitem{lihb}
H.~B.~Li and X.~R.~Lyu,
\href{https://doi.org/10.1093/nsr/nwab181} {Natl. Sci. Rev. \textbf{8}, nwab181 (2021).}

\end{thebibliography}
\end{document}